\documentclass[aps,twocolumn,superscriptaddress,altaffilletter,lengthcheck,tightenlines]{revtex4}

\usepackage{multirow}

\newcommand{\ben}{\begin{eqnarray}}
\newcommand{\een}{\end{eqnarray}}
\newcommand{\be}{\begin{equation}}
\newcommand{\ee}{\end{equation}}
\newcommand{\ba}{\begin{eqnarray}}
\newcommand{\ea}{\end{eqnarray}}
\newcommand{\n}{\label}
\newcommand{\no}{\noindent}

\newcommand{\ro}{\rho}

\usepackage[dvipdf]{epsfig}
\usepackage{color}

\begin{document}
\date{\today}

%%%%%%%%%%%%%%%%%%%%%%%%%%%%%%%%%%%%%%%%%%%%%%%%%%%%%%%%%%%%%%%%%%%%%%%%%%%%%%%%%%%%%%%%%%%%%%%%%%%%%%%%%%%%

\title{Comments on \emph{Unified dark energy and dark matter from a scalar field different from quintessence}}

%%%%%%%%%%%%%%%%%%%%%%%%%%%%%%%%%%%%%%%%%%%%%%%%%%%%%%%%%%%%%%%%%%%%%%%%%%%%%%%%%%%%%%%%%%%%%%%%%%%%%%%%%%%%

\author{Luis P. Chimento}\email{chimento@df.uba.ar}
\affiliation{ Departamento de F\'{\i}sica, Facultad de Ciencias Exactas y
Naturales,  Universidad de Buenos Aires, Ciudad
Universitaria, Pabell\'on I, 1428 Buenos Aires, Argentina}

\author{M\'onica  Forte}\email{forte.monica@gmail.com}
\affiliation{ Departamento de F\'{\i}sica, Facultad de Ciencias Exactas y
Naturales,  Universidad de Buenos Aires, Ciudad
Universitaria, Pabell\'on I, 1428 Buenos Aires, Argentina}

\begin{abstract}

In a recent paper by C. Gao, M. Kunz, A. Liddle and D. Parkinson \cite{Gao:2009me} the unification of dark matter and dark energy was explored within a theory containing a scalar field of non-Lagrangian type. This scalar field, different from the classic quintessence, can be obtained from the scalar field representation of an interacting two-fluid mixture described in the paper by L.P. Chimento and M. Forte \cite{Chimento:2007da}. 

\end{abstract}

%\pacs{}
%\keywords{dark energy theory}

\maketitle

\section{Introduction}

Due to the additivity of the stress-energy tensor is possible to describe the quintessence field in terms of two interacting fluids, namely, stiff matter and vacuum energy. Besides, the total energy-momentum conservation is equivalent to the Klein-Gordon equation \cite{Chimento:2002gb}-
\cite{Chimento:2005wv}. This description was generalized assuming a mixture of two fluids with constant equations of state, which interact between them in a flat Friedmann-Robertson-Walker model \cite{Chimento:2007da}. There, a scalar field $\phi$ representation of the mixture of two interacting fluids was introduced by imposing the condition that $\dot\phi^2$ is an appropriate linear combination of the energy densities $\ro_1$ and $\ro_2$ of both fluids. Under these conditions, the resulting model was called exotic quintessence.

%%%%%%%%%%%%%%%%%%%%%%%

\section{Exotic quintessence}
%%%%%%%%%%%%%%%%%%%%%%%%%%%%%%%

We present the exotic quintessence model developed in \cite{Chimento:2007da} including two extra matter components, with energy densities $\ro_3$ and $\ro_4$, which separately satisfy their own equations of conservation. In this case, the Einstein equations read:
\ben
\n{00}
3 H^2 =\rho_1 +\rho_2+\rho_3+\rho_4,\\
\n{c}
\dot\rho_1+\dot \rho_2 + 3H[(1+w_1)\rho_1 + (1+w_2)\rho_2]= 0,\\
\dot\rho_3+ 3H(1+w_3)\rho_3=0,\\
\dot\rho_4+ 3H(1+w_4)\rho_4=0,
\een

\no where we have adopted constant equations of state $w_n=p_{n}/\rho_{n}$ for each fluid with $n$ = 1,2,3,4. Units are chosen so that the gravitational constant is set to $8\pi G=1$ and $c=1$. We introduce a scalar field $\phi$ representation of the interacting two first fluids associating $\dot\phi^2$ with the following linear combination of $\ro_1$ and $\ro_2$ 
\be
\n{f2}
\dot\phi^2=(1+w_1)\rho_1+(1+w_2)\rho_2.
\ee

From Eqs. (\ref{c}) and (\ref{f2}) we obtain the total energy density and pressure of the four-fluid mixture, and the dynamical equation for the scalar field 
\ben
\n{r}
\rho=\frac{\dot\phi^2}{1+w_1}+\frac{w_1-w_2}{1+w_1}\,\,\rho_2+\rho_3+\rho_4,\\
p=\frac{w_1\dot\phi^2}{1+w_1}-\frac{w_1-w_2}{1+w_1}\,\,\rho_2+w_3\rho_3+w_4\rho_4,\\
\n{kg}
\ddot\phi + \frac{3}{2}(1+w_1) H\dot\phi+ \frac{w_1 - w_2}{2}\,\frac{\dot\rho_2}{\dot\phi} = 0.
\een

These equations define the exotic quintessence model.

\section{Conclusions}

\no Choosing $w_1=0$, $w_3=0$, $w_4=1/3$, and making the substitutions $\phi \to \phi/\sqrt{2}$ and $-w_2\rho_2\to \Lambda(\phi)$, the Einstein and the exotic quintessence equations are reduced to the expressions (21)-(23) of the paper \cite{Gao:2009me}. On the other hand, if we impose the integrability condition mentioned in \cite{Chimento:2007da} 
\be
\n{ci}
\dot\ro_2+A\dot\phi\ro_2=0,
\ee
on the field equation (\ref{kg}) 
with a constant $A$, we get $\ro_2\propto \Lambda(\phi)\propto \exp{(-A\phi)}$ after integrate the Eq. (\ref{ci}). Then, the exotic quintessence model with the integrability condition (\ref{ci}) leads to an exotic scalar field driven by an exponential potential and thus to obtain the model investigated in Ref. \cite{Gao:2009me}. As was already pointed in Ref. \cite{Chimento:2007da}, this model is integrable and can be solved exactly.

\end{document}